\documentclass[conference]{IEEEtran}
\IEEEoverridecommandlockouts
\usepackage{cite}
\usepackage{url}
\usepackage{amsmath,amsthm,amssymb,amsfonts}
\usepackage{algorithmic}
\usepackage{graphicx}
\usepackage{textcomp}
\usepackage{xcolor}
\usepackage{comment}
\usepackage[hidelinks]{hyperref}

\usepackage{graphicx} 
\usepackage{float} 
\usepackage{subfigure} 

\def\BibTeX{{\rm B\kern-.05em{\sc i\kern-.025em b}\kern-.08em
    T\kern-.1667em\lower.7ex\hbox{E}\kern-.125emX}}

\IEEEoverridecommandlockouts
\IEEEpubid{\makebox[\columnwidth]{978-1-6654-6934-0/22/\$31.00~\copyright~2022 IEEE \hfill}\hspace{\columnsep}\makebox[\columnwidth]{ }}

\begin{document}

\title{Driver-Side and Traffic-Based Evaluation Model for On-street Parking Solutions
}

\author{
\IEEEauthorblockN{Qianyu Ou$^1$, Wenjun Zheng$^2$, Zhan Shi$^1$ and Ruizhi Liao$^3$$^,$$^4$}
\IEEEauthorblockA{$^1$School of Data Science, The Chinese University of Hong Kong, Shenzhen, China \\ $^2$School of Science and Engineering, The Chinese University of Hong Kong, Shenzhen, China \\ $^3$School of Humanities and Social Science, The Chinese University of Hong Kong, Shenzhen, China \\  $^4$Shenzhen Key Laboratory of IoT Intelligent Systems and Wireless Network Technology, China \\
\{qianyuou, wenjunzheng, zhanshi1\}@link.cuhk.edu.cn, rzliao@cuhk.edu.cn}}

\maketitle

\begin{abstract}
Parking has been a painful problem for urban drivers. The parking pain exacerbates as more people tend to live in cities in the context of global urbanization. Thus, it is demanding to find a solution to mitigate drivers' parking headaches. Many solutions tried to resolve the parking issue by predicting parking occupancy. Their focuses were on the accuracy of the theoretical side but lacked a standardized model to evaluate these proposals in practice. This paper develops a Driver-Side and Traffic-Based Evaluation Model (DSTBM), which provides a general evaluation scheme for different parking solutions. Two common parking detection methods - fixed sensing and mobile sensing - are analyzed using DSTBM. The results indicate: first, DSTBM examines different solutions from the driver's perspective and has no conflicts with other evaluation schemes; second, DSTBM confirms that fixed sensing performs better than mobile sensing in terms of prediction accuracy.
\end{abstract}


\section{Introduction}
It is not only a pain, but also a waste (e.g., time or fuel) for drivers to search for parking spaces in cities. It is estimated that up to 7 billion people, meaning more than two-thirds of the world population, will live in urban areas by 2050 \cite{b1}. The growing urbanization rate leads to higher car ownership that aggravates the urban parking pain. As cities only have limited land spaces, urbanization and parking form a group of conflicts competing for land resources. Thus, it is essential to find effective solutions to fix the parking problem \cite{b2}. 

According to the reports \cite{b3,b4}, West Europe has around 300 million parking spots, while the United States has 2 billion ones. Many of those parking spaces are not used effectively due to a lack of real-time information on parking occupancy. Thus, smart parking can play a vital role in mitigating the urban parking pain by bringing in instant parking information to make the best use of city parking facilities.

\begin{figure*}[htbp] 
\centering 
\includegraphics[width=0.75\textwidth]{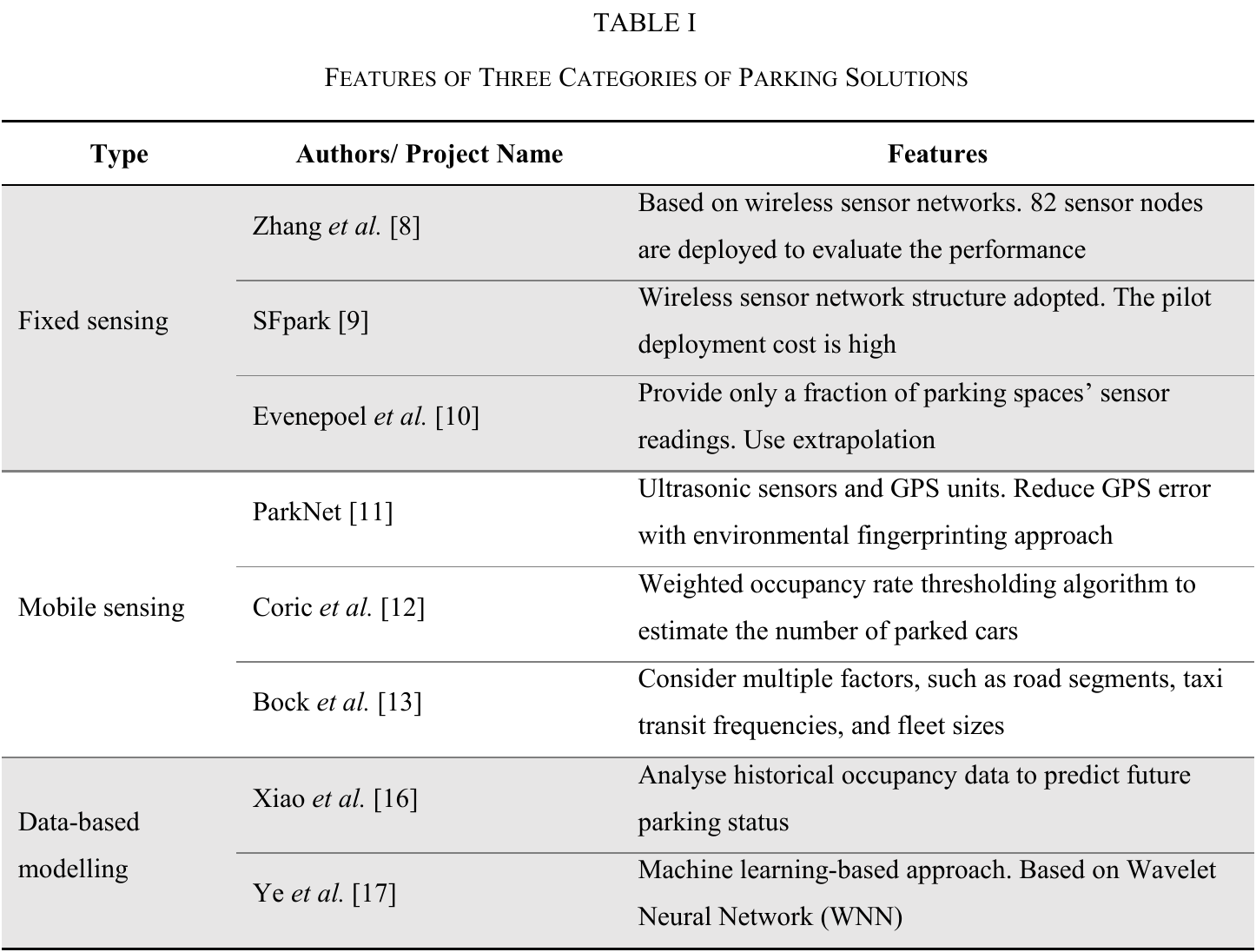} 
\end{figure*}

In general, parking solutions can be divided into three categories: fixed sensing, mobile sensing, and data-based modeling \cite{b5,c1}. Conventionally, fixed sensing is the most used one among the three solutions, the mechanism of which is to install sensors at the parking spaces to detect the parking occupancy. As for mobile sensing, sensors are attached to moving vehicles and take the chance when vehicles pass by the parking region to gain the parking information. In mobile sensing, one sensor keeps an eye on multiple parking spaces, reducing the cost for sensor implementations compared with the fixed sensing solution. In the past years, data-based modeling has become a popular tool as the advancement of machine learning, which utilizes historical data to estimate the future parking occupancy.

An evaluation model is needed to verify the proposed parking solution. This paper presents a Driver-Side and Traffic-Based Evaluation Model (DSTBM) for different parking solutions from a new angle - the driver's perspective. Usually, the conventional evaluation models focus on the accuracy of parking occupancy. In DSTBM, we use traffic flows to set up our model, and predict the parking status from the driver's perspective, which means the driver's decision (to park or not to park) is also included into the accuracy calculation of DSTBM.

DSTBM consists of two parts: a driver's decision model and a simulation process. In the driver's decision model, we set a group of policies to help drivers decide whether to park or not. The decision results and traffic data are taken as the inputs for the simulation process, which simulates the drivers' parking behaviors and estimates the parking status.

The rest of the paper is organized as follows: section \uppercase\expandafter{\romannumeral2} reviews evaluation models in the literature, summarizes  their features, and introduces the main contribution of our work; section \uppercase\expandafter{\romannumeral3} presents DSTBM by explaining assumptions and the simulation mechanism; section \uppercase\expandafter{\romannumeral4} shows the performance evaluation of fixed sensing and mobile sensing solutions proposed in \cite{b7} using DSTBM; section \uppercase\expandafter{\romannumeral5} concludes the paper and looks into the future work.

\section{Related work}
This section reviews three types of parking solutions: fixed sensing \cite{b8,b9,b10}, mobile sensing \cite{b11,b12,b13,b14,b16}, and data-based modeling \cite{b15,b6}, whose features are summarized in Table \uppercase\expandafter{\romannumeral1}.

Zhang \emph{et al.} \cite{b8} present a street parking system (SPS) based on wireless sensor networks. This system equips magnetic sensor nodes on parking spots to detect parking occupancy. In total 82 sensor nodes are installed to evaluate the performance of SPS. The authors claim SPS can achieve a detection accuracy of around 98\%.  

SFpark \cite{b9} is a U.S. parking project started by San Francisco Municipal Transportation Agency (SFMTA). Eight thousand parking spaces were equipped with 11711 magnetometer sensors. Parking availability data collected by parking sensors were periodically broadcasted, so that drivers can save time on cruising for parking spaces and congestion can be reduced.

One of the shortcomings of fixed sensing solutions is the high deployment cost. To lower the costs, Evenepoel \emph{et al.} \cite{b10} utilize sensor information on a portion of parking spaces to estimate the city-wide parking status using extrapolation. In addition, they build a probabilistic model to calculate the degree of dependence between the performance of sensor networks and the number of installed sensors. They define the performance measure based on the use of "redirection threshold". Due to the computational complexity of the model, the estimated number of discrete points does not always match the reality.

For mobile sensing solutions, there is no need to equip individual parking spots with sensors. ParkNet \cite{b11} aims to build a real-time parking map. The authors employ ultrasonic sensors and GPS units to determine the number of parked vehicles. In addition, they devise an environmental fingerprinting to improve the precision of GPS. The authors took a two-month drive test to collect 500-mile parking data and claimed the accuracy of the parking occupancy map is over 90\%.

Coric \emph{et al.} \cite{b12} build a parking map using pre-installed parking sensors on vehicles. Unlike ParkNet, the authors propose a weighted occupancy rate threshold algorithm to estimate the number of vehicles parked on the streets. They collect more than 2 million sensor readings and demonstrate that the accuracy of their parking maps is around 90\%.

Bock \emph{et al.} \cite{b13} simulate the sensing coverage of on-street parking by down-sampling the parking data from SFpark. They assume that a fleet of taxis equipped with sensors are able to detect free on-street parking spots. According to the taxi trajectories, the authors estimate the sensing coverage of probing taxis with different number of taxis. Bock \emph{et al.} \cite{b14} and Liao \emph{et al.} \cite{b16} make a further investigation on the suitability of taxis to crowdsense on-street parking availability. Multiple factors are considered, such as road segments, taxi transit frequencies, and fleet size. The results show that crowd-sensing parking occupancy via taxis is a promising alternative to the expensive fixed-sensing solution.

\begin{figure*}[htbp] 
\centering 
\includegraphics[width=0.75\textwidth]{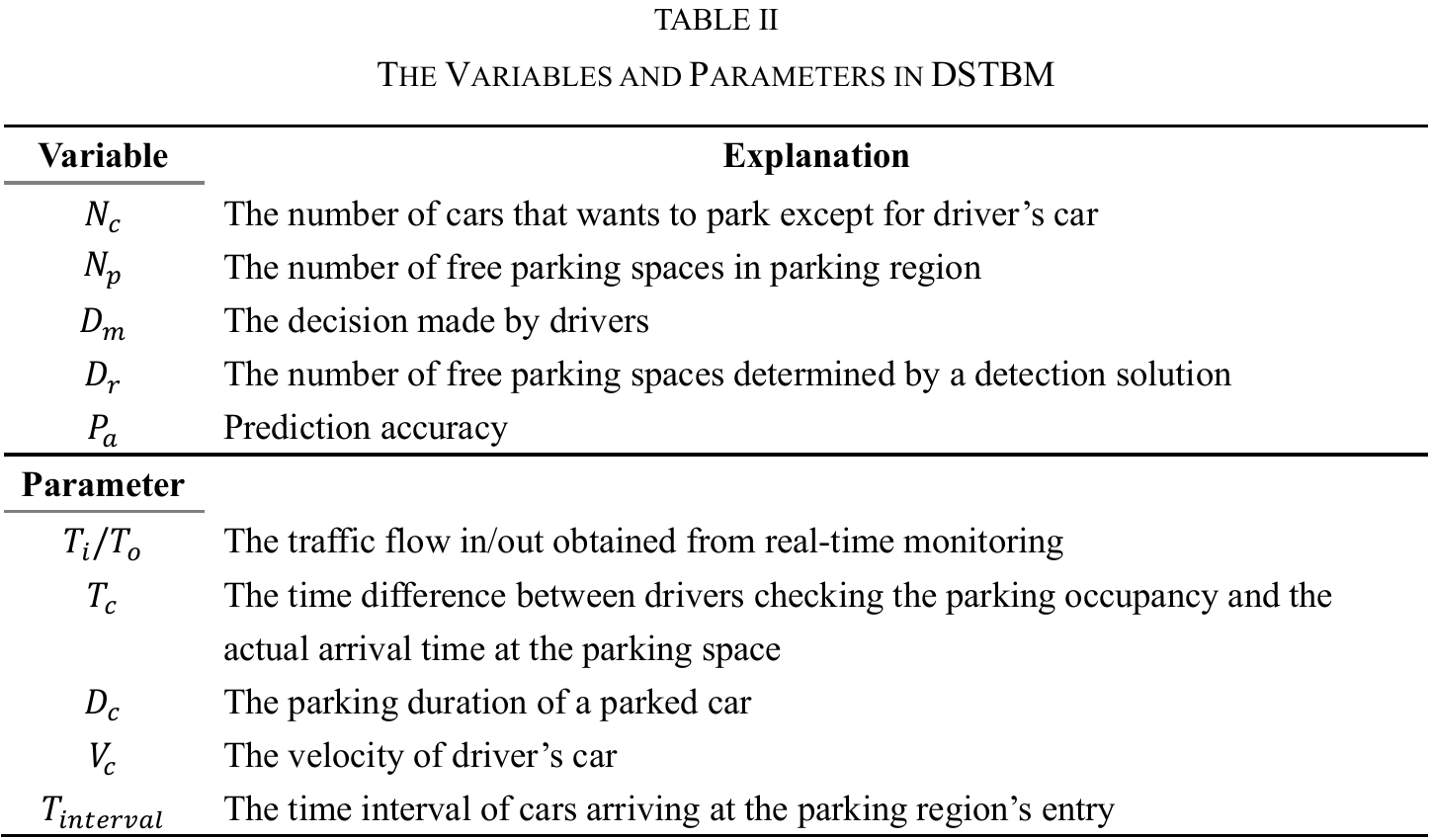} 
\end{figure*}

Data-based modeling makes use of data such as traffic conditions, historical parking data, and drivers' parking demands. Xiao \emph{et al.} \cite{b15} put forward a model-based practical framework, the core idea of which is to utilize a queuing model to analyze the historical occupancy data and then predict the future parking status. The authors applied their model to study parking spaces in San Francisco. The results indicate that their model can predict parking occupancy with good accuracy. 

Ye \emph{et al.} \cite{b6} proposed a machine learning-based approach to make a short-term estimation of available parking spaces (APS). The prediction model is based on Wavelet Neural Network. The authors show the positive impacts of APS, such as lessening cruising time, reducing illegal parking behaviors, and providing information for adjusting parking rates dynamically. The prediction accuracy is compared with that of a control group.

Zheng \emph{et al.} in \cite{b17}  used historical parking data to evaluate the performance of a mobile sensing solution. The analytical results are compared with the ground truth data. However, the model only takes the number of parking spaces into the calculation, which is too simplified.

According to the reported results in \cite{b9}, the accuracy of SFpark, a complete fixed sensing solution, is around 86\%, while some mobile sensing solutions' accuracy can be above 90\%. The common evaluation method used for mobile sensing is to check the prediction result against the ground truth to decide the detection accuracy. However, the works in the literature ignore that there is a time gap between the detected parking occupancy and the actual arrival time of drivers. For example, when drivers receive the parking occupancy updates, they may still be several blocks away from the preferred parking spot. It implies that we should take drivers' decisions when drivers arrive at the preferred parking region into the parking evaluation model. Thus, the main contribution of this paper is that we present a unified evaluation model integrating drivers' decisions with other factors (e.g., traffic conditions) for different parking detection solutions.

\section{Driver-Side and Traffic-Based Evaluation Model}
In this section, we develop a Driver-Side and Traffic-Based Evaluation Model (DSTBM) aiming to evaluate parking solutions by incorporating drivers' perspectives. DSTBM consists of two parts: a driver's decision model and a simulation process. The driver's decision model generates the driver's parking decision based on real-time traffic conditions and driver's consideration. The simulation process takes drivers' decisions derived from the decision model as inputs, and then gives the prediction accuracy of the concerned parking detection solution. The corresponding algorithms are shown in Algorithm1 and Algorithm2 in Appendix.


\subsection{DSTBM Preliminary}
We build DSTBM to provide a standardized evaluation method for parking detection. One fact and four assumptions of the model are explained as follows. The variables and parameters of the model are shown in Table \uppercase\expandafter{\romannumeral2}. 

\begin{itemize}
\item Fact 1: Every car has a GPS, and we can obtain the car's position and velocity from GPS. 
\item Assumption 1: The velocity of a car that goes through a parking region is considered to be constant. Namely, a car goes through a parking lot containing multiple parking spots without changing its speed.
\item Assumption 2: A single lane inside the parking region, or a single queue, is assumed. Namely, a car willing to park will not leave the parking region before a car ahead of it has parked or exited the parking region.
\item Assumption 3: The parking duration of a car stayed in a parking spot ($D_c$) is a continuous random variable and follows the normal distribution with mean $\mu$ and variance $\sigma^2$, both of which can be determined by historical parking data.  
\item Assumption 4: The time interval of cars arriving at the parking region ($T_{interval}$) is a continuous random variable and follows the exponential distribution with an arriving rate $\lambda$. $T_{interval}$ is memoryless, meaning the intervals of the following drivers are unrelated.
\end{itemize}

\begin{figure*}[htbp] 
\centering 
\includegraphics[width=0.75\textwidth]{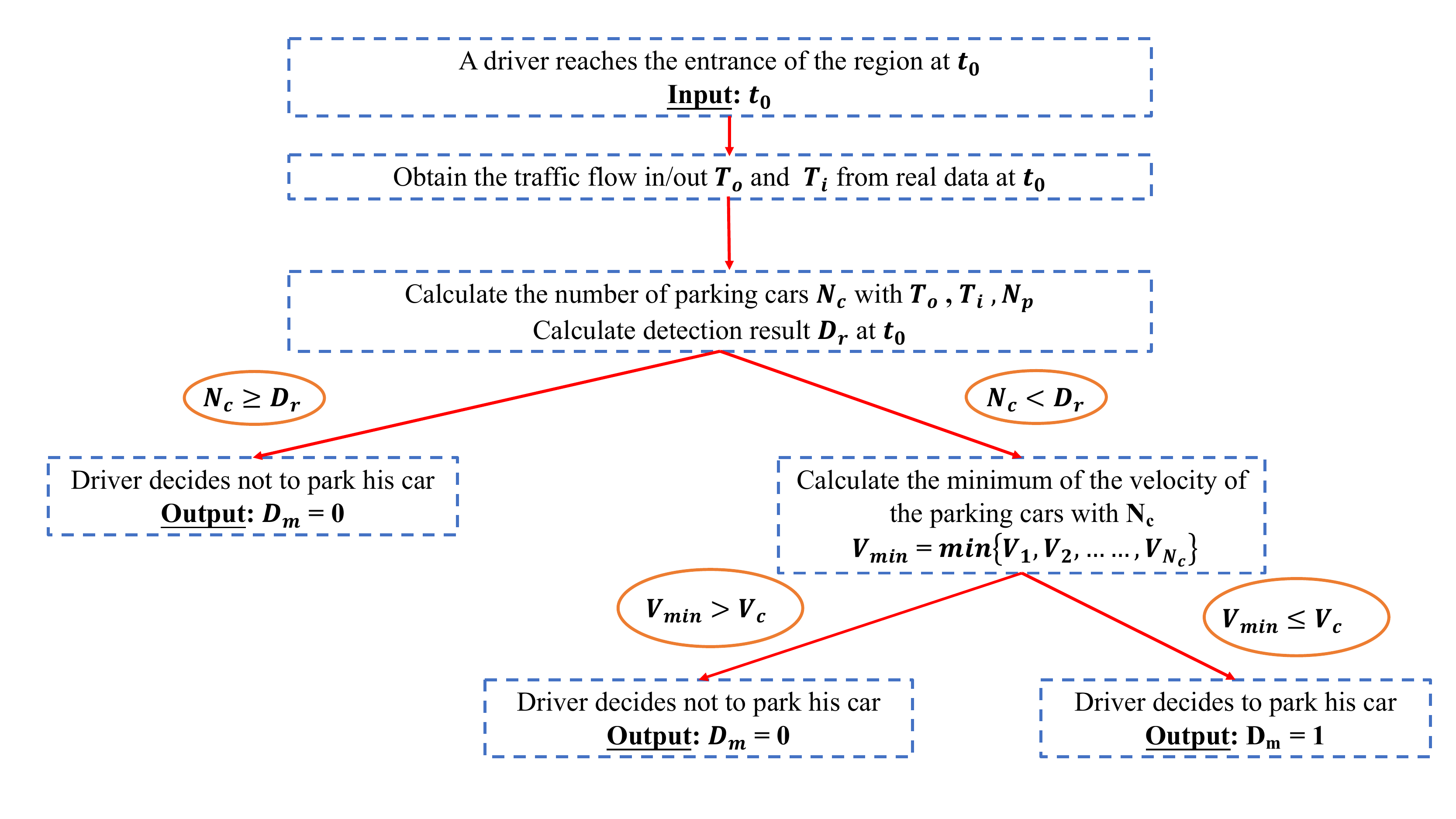} 
\caption{Driver's Decision Model} 
\end{figure*}

\subsection{Driver's decision model}
As mentioned above, in a real driving practice, there will likely be  a time gap between the detected parking availability and the actual arrival time of the driver. In order to make driver's decision closer to reality, we develop a driver's decision model by introducing this time gap. The core idea is to jointly utilize other influential factors (e.g., traffic conditions) and the detection result based on crowdsourcing to predict a driver's decision in the real parking situation.

The mechanism of the driver's decision model is illustrated in Fig. 1. The starting point of the model is $t_0$, and the output is $D_m$, i.e., the parking decision that the driver makes. When a driver reaches the entrance of a parking region, the model is activated. The number of cars ($N_c$) that are looking for parking spaces except for driver's car is calculated based on the traffic flow data. There are three cases for drivers, as $N_c$ and $D_r$ (detection result of parking spaces) form different combinations. If $N_c \geq D_r$, it indicates there are not enough parking spaces, so the driver will decide not to park ($D_m=0$). If $N_c<D_r$, the driver is likely to obtain a parking space but not guaranteed because $D_r$ may not be completely accurate. So, the driver will consider further. If $V_{min}> V_c$, it means the driver's car is slower than all other cars. Once the $D_r$ is not accurate, the driver will fail to compete with others due to the car's low speed, so $D_m=0$ as well. If $N_c<D_r$ and $V_{min}\leq V_c$, that is the driver is not slower than all those looking for parking spaces, then the driver can arrive at the parking space earlier than some other cars, so the driver will decide to park in the region ($D_m=1$).

\begin{figure*}[htbp] 
\centering 
\includegraphics[width=0.85\textwidth]{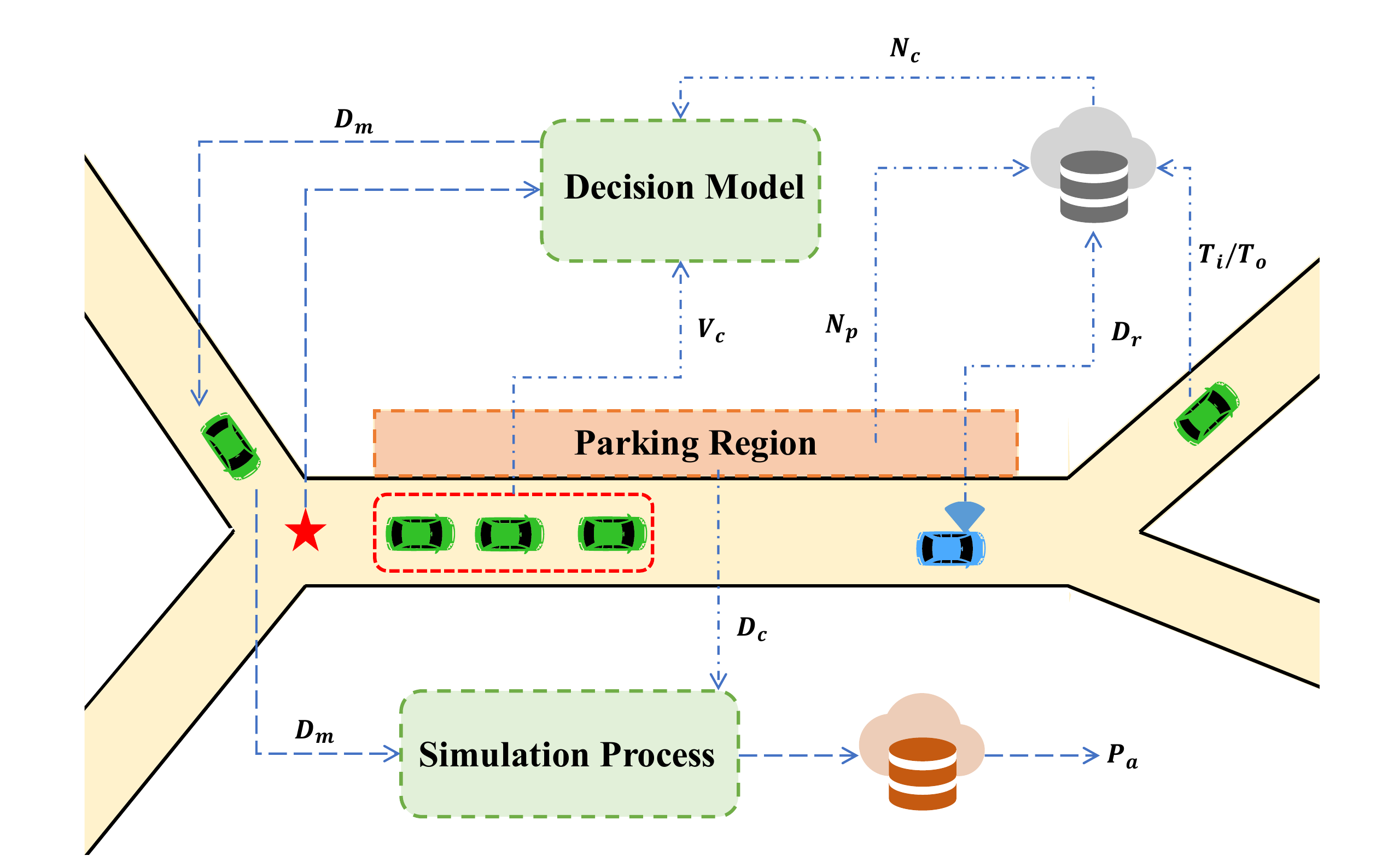} 
\caption{Driver-Side and Traffic-Based Model} 
\label{Fig.2} 
\end{figure*}
\subsection{Simulation Process}
In the simulation process, we will trace the driver's behavior after they have made their decisions to obtain the prediction accuracy $P_{a}$ of a parking solution. 
As explained above, the driver's decision model has two outcomes: $D_m=0$ or $D_m=1$. The former means drivers decide not to park based on parking detection results and traffic conditions (e.g., the number of cars looking for parking spaces). If $D_m=0$, but there are actually chances to park, it is a false negative, namely, a wrong prediction. The latter means drivers decide to park. If $D_m=1$, but drivers fail to park finally, it is a false positive, which is also a wrong prediction. The drivers' actual parking behaviors can be examined when the cars finally arrive at the parking spaces (i.e., after $T_c$) by simulations. Thus, the prediction accuracy $P_{a}$ of a parking solution can be calculated as true positive plus true negative:
\begin{align*}
P_{a}=&\frac{free\,space\,predicted\,\&\,driver\,actually\,park}{total\,prediction} + \\
&\frac{no\,free\,space\,predicted\,\&\,driver\,actually\,not\,park}{total\,prediction}. 
\end{align*}



The simulation codes have been made available via the link below.\footnote{\url{https://www.dropbox.com/s/0rcunfnunh2p6f9/Driver_Side_and_Traffic_Based_Evaluation_Model.ipynb?dl=0}}

\subsection{DSTBM Summary}
As shown in Fig. 2, the car in blue represents a randomly moving vehicle that is equipped with sensors to detect on-street parking occupancy, while the car in green represents a normal car. 

Note that, the number of free parking spaces $N_p$, detection result $D_r$, and traffic flow data $T_i/T_o$ are variables, which will be continuously updated to the data processing center to compute $N_c$. According to \emph{Assumption 4}, the interval ($T_{interval}$) between adjacent cars arriving at the parking region follows exponential distribution with a coming rate of $\lambda$. If we denote $G (T_{interval})$ as the probability density function of the arrival interval, then, the probability that adjacent cars enter the parking region is equal to $\int^{t+\delta_1}_{t-\delta_1}G(x)dx$, where $\delta_1$ is a small number and $G(x) = \lambda e^{-\lambda x},\,x \geq 0$. 

When a car arrives at the entrance of the parking region (the red star in Fig. 2), the decision model is activated, which combines $N_c$ and $V_c$ to produce an output ($D_m$): whether to park or not. Drivers will then behave according to the output of the decision model. Meanwhile, $D_m$ will be transmitted as inputs to the simulation process. $D_m$ plus $D_c$ (the parking duration of a car) combined are used to determine the prediction accuracy. According to \emph{Assumption 3}, $D_c$ follows normal distribution. So, the probability that the car still parks in the parking spot at the next moment is equal to $\int^{d+\delta_2}_{d-\delta_2}F(y)dy$, where $\delta_2$ is a small number and $F(y)=\frac{1}{\sigma\sqrt{2\pi}}e^{-\frac{(y-\mu)^2}{2\sigma^2}}, y \geq 0$.

\section{Simulation Results}

With the help of DSTBM, we can evaluate the prediction accuracy of different parking solutions. To this end, we make two comparisons, as shown in Fig. 3 and Fig. 4. Both figures take the time (week) as the abscissa, while the ordinate represents the detection accuracy. In the two figures, the lines in different colors represent different detection schedules ($D_s$). The detection schedule means the interval that a crowdsourcing detection car scans a specific parking spot. For example, in Fig. 3, $D_s=0$ means a fixed sensing solution that constantly updates the parking status, while $D_s=15$ means that a detecting car passes by a parking spot and updates the parking status every 15 minutes.

\begin{figure}[htbp] 
\includegraphics[width=0.48\textwidth]{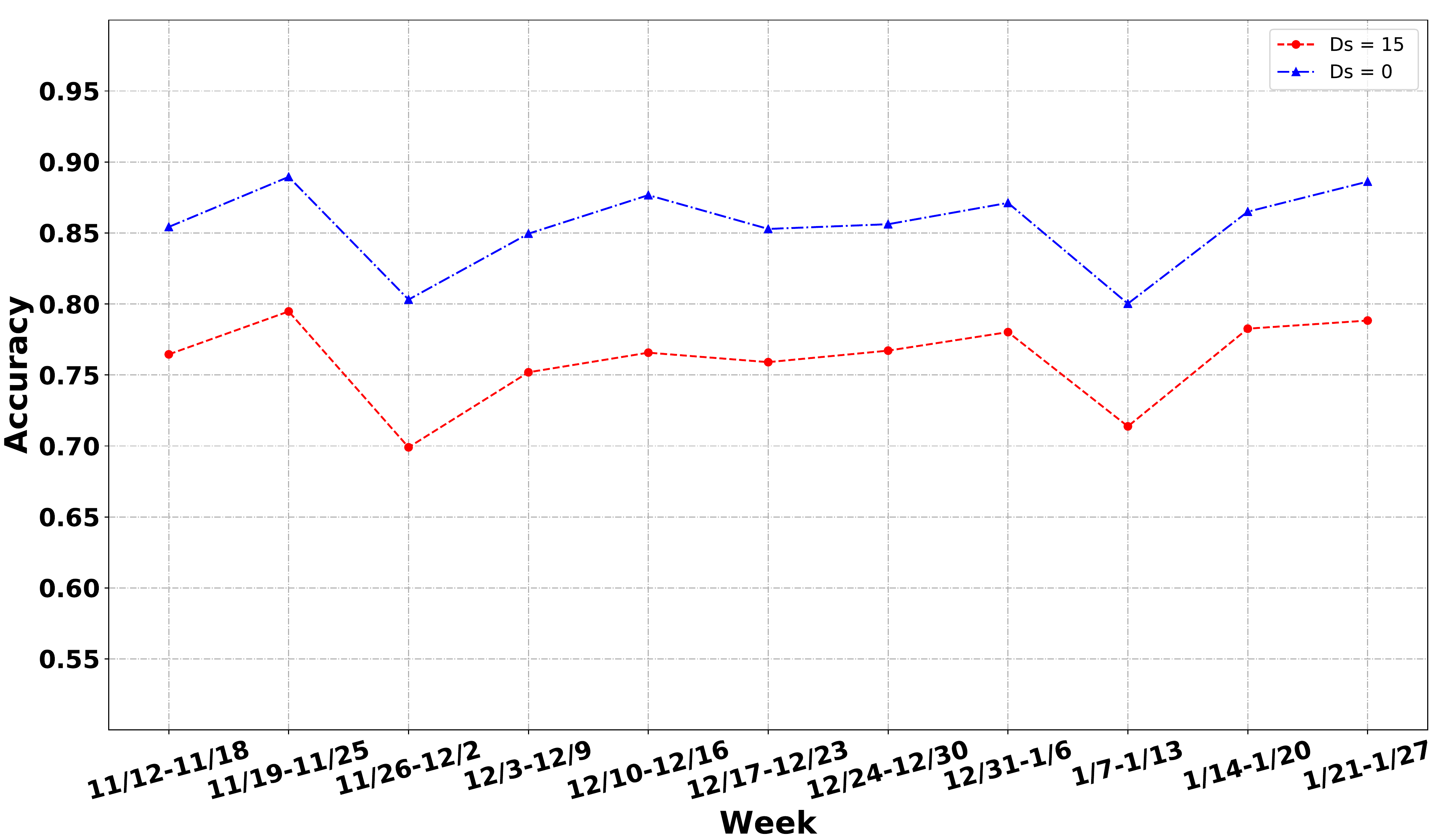} 
\caption{Fixed sensing vs mobile sensing} 
\label{Fig.3} 
\end{figure}

\begin{figure}[htbp] 
\includegraphics[width=0.48\textwidth]{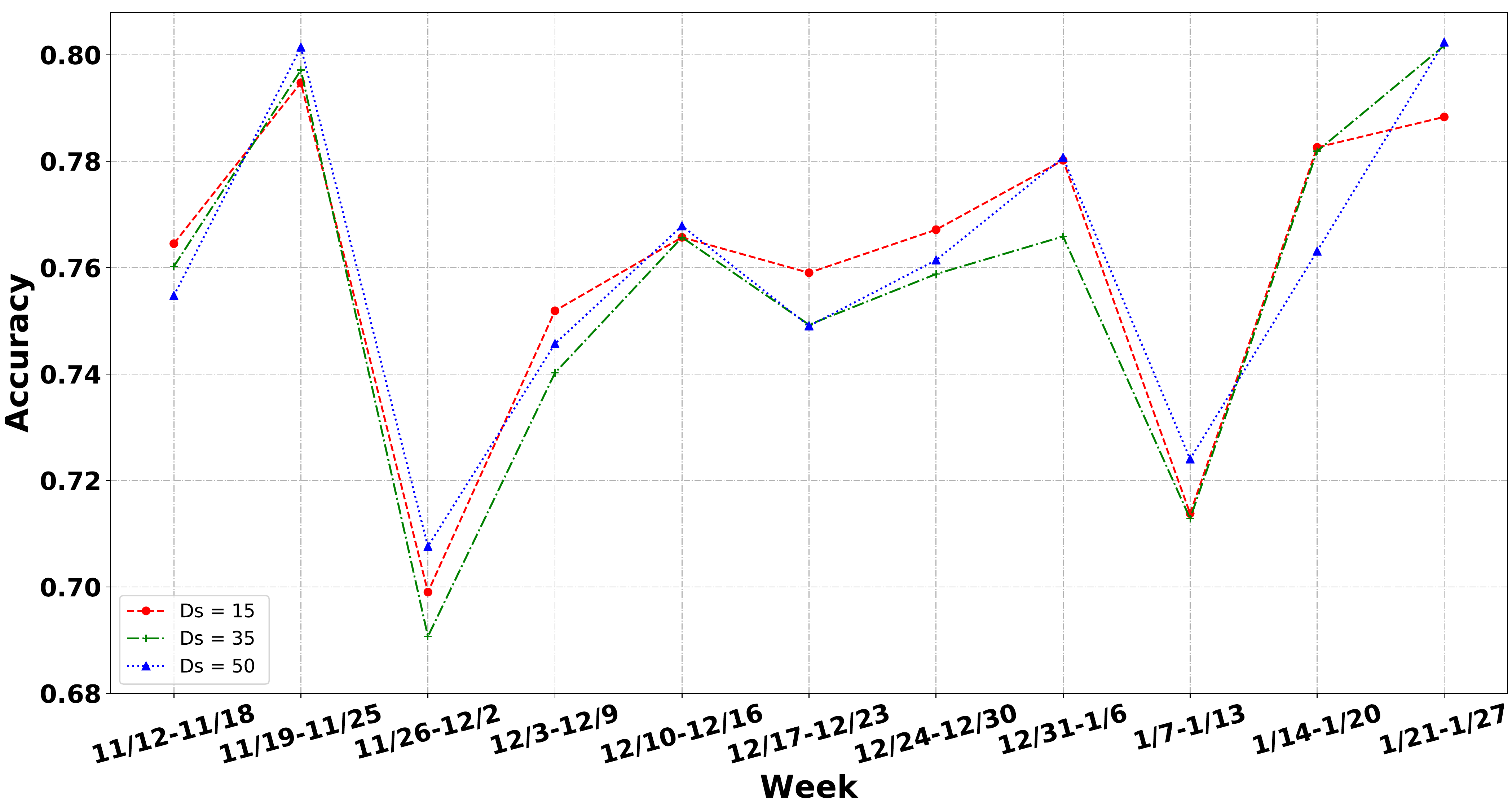} 
\caption{Mobile sensing with different detection schedules} 
\label{Fig.4} 
\end{figure}

\subsection{Comparing fixed sensing and mobile sensing}

We collected a set of real on-street parking occupancy data in Nanshan, Shenzhen from 2018/11/12 to 2019/01/27 \cite{b18}. We use this data set as the result of the fixed sensing solution, since the detection schedule is $D_s=0$.

Also, we used this data set to simulate the detection result given by a fleet of crowdsensing vehicles, which are assumed to be equipped with detection sensors, periodically cruising the Nanshan streets, and passing by on-street parking spaces every $D_s$. 

From the results shown in Fig. 3, the accuracy of the fixed sensing solution is around 90\%, which is consistent with the situation in practice. The accuracy of the simulated mobile sensing solution is not as good as that of fixed sensing, which fluctuates around 80\%. This is evident that DSTBM has no conflicts with the other evaluation models. It is also reasonable that the accuracy of mobile sensing is lower than that of fixed sensing, as mobile sensing sacrifices some accuracy in pursuit of reducing the number of sensor deployment costs.

\subsection{Mobile sensing with different detection schedules}
We can simulate the mobile sensing solution by setting different detection schedules to process real parking occupancy data in Nanshan. For instance, suppose that the detection schedule is every 10 minutes ($D_s=10$), and let 0 and 1 denote the parking status (0 for free, and 1 for occupied). If the initial parking status is occupied, and the sensing vehicle scans the parking spaces in the first minute of every 10 minutes, then, the parking status gained by simulated mobile sensing is 1111111111. But in a busy street, the actual parking status from the minute 1 to the minute 10 can be 1110010001. So, we will get different detection results by processing the real data set using different $D_s$. If $D_s = 0$, it is equivalent to fixed sensing.

The mobile sensing solution is simulated assuming the taxi fleet is equipped with detection sensors with different detection schedules: 15min, 35min and 50min.

In general, the accuracy is higher when the detection schedule is smaller, indicating the accuracy of parking solutions decreases as the detection interval increases. As shown in Fig. 4, the accuracy when $D_s=15$ (red line) is generally higher than other lines. However, there are some cases that the accuracy of $D_s=50$ is higher than that of $D_s=15$. These occurrences are due to the fact that our simulation is based on real parking data (but a relatively small set), so that fluctuations are inevitable. The other reason is that we derive the model by adding the driver's perspective, which can lead to different results from fixed sensing.


\subsection{Discussion}
Results of the above two sub-sections all match the situations in practice. The DSTBM model examines detecting solutions from both the driver's perspective and real-time traffic conditions, thus this reality-based evaluation model conforms with the results of other evaluation models. In addition, the DSTBM model considers more practical factors, making it a more robust evaluation method. 

\section{Conclusions}
In this paper, we proposed a Driver-Side and Traffic-Based Evaluation Model (DSTBM) aiming to recover the lack of a unified evaluation model for different parking solutions. The DSTBM model is consisted of two parts: a driver's decision model and a simulation process, which simulates different parking solutions and outputs the detection accuracy. The driver's decision model generates reasonable suggestions based on different traffic parameters. Then, based on those suggestions, the simulation model gives the prediction accuracy by incorporating the driver's perspective. 

The fixed sensing solution and mobile crowdsourcing solution proposed in \cite{b7} are taken to assess the performance of DSTBM. DSTBM maintains the same conclusion: fixed sensing solutions have a higher prediction accuracy than the mobile sensing ones. In addition, DSTBM introduces many practical factors (e.g., traffic flow and drivers' arrival time), making the model more robust.

For the future work, we will consider a more general case, in which drivers randomly check parking information until arriving at the parking region. To fully capture this feature, we will adopt the dynamic programming in the driver's decision model.

\section*{Acknowledgement}
This work was supported in part by NSFC (61902332), Shenzhen STIC (JCYJ20180508162604311), Longgang STIB (20200030) and CUHK(SZ) URA.

\begin{appendices}
\section{DSTBM Algorithms}
\begin{figure}[H] 
\centering 
\includegraphics[width=0.39\textwidth]{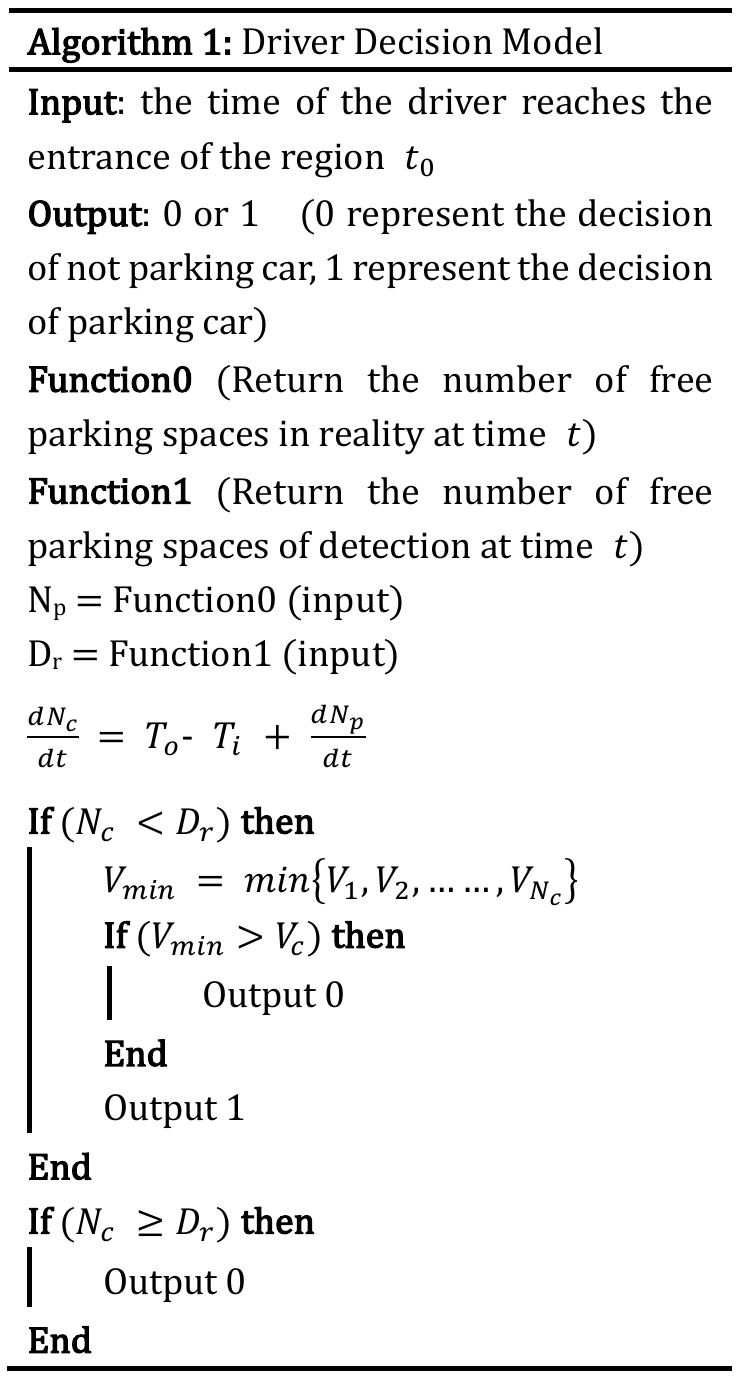} 
\label{Algorithm1} 
\end{figure}

\begin{figure}[H] 
\centering 
\includegraphics[width=0.4\textwidth]{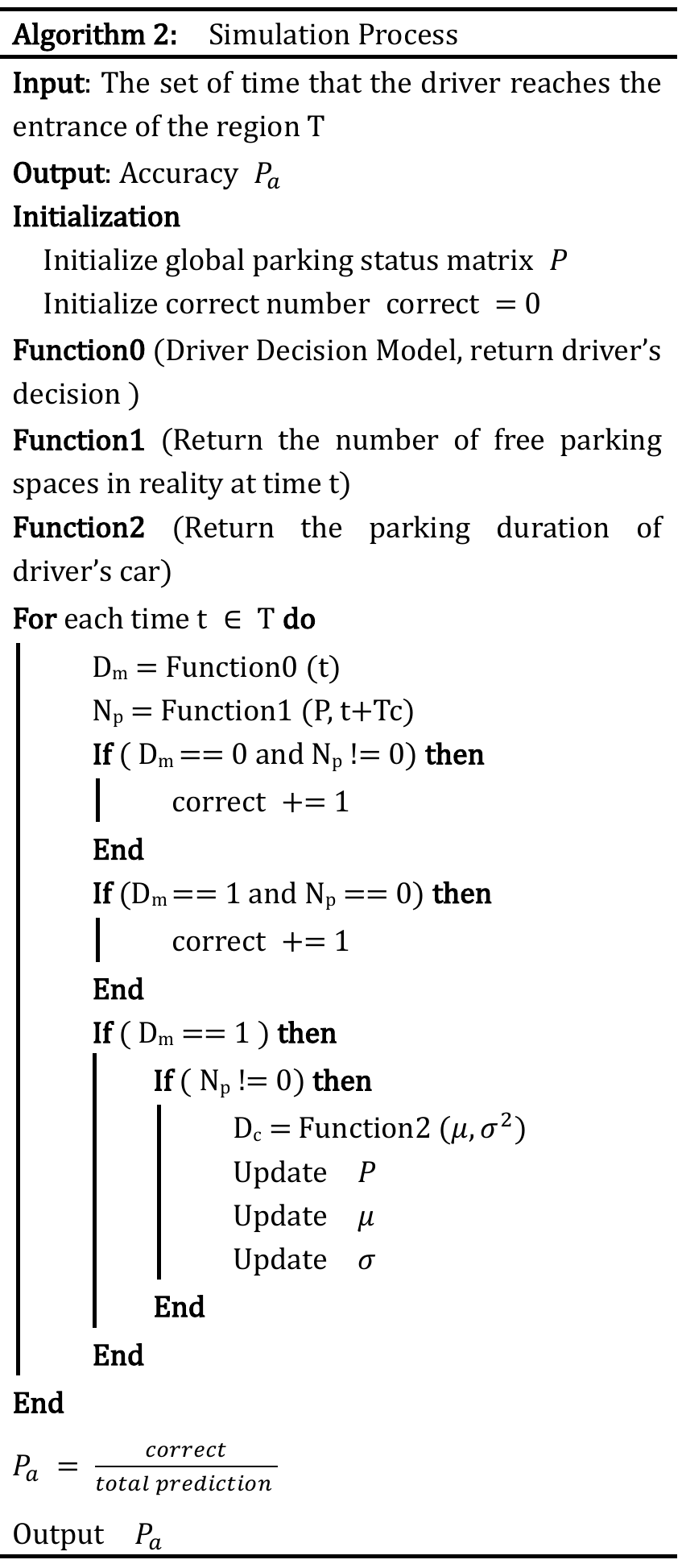} 
\label{Algorithm2} 
\end{figure}
\end{appendices}

\end{document}